\documentclass[preprint]{revtex4-1}

\usepackage{dcolumn}
\usepackage{amsbsy}
\usepackage{amsmath}
\usepackage{fullpage}
\usepackage{subfigure}
\usepackage{amssymb}
\usepackage{color}
\usepackage{graphicx}
\usepackage{epstopdf}
\usepackage{bm}
\usepackage{caption}
\usepackage{cleveref}
\usepackage{mathrsfs}
\usepackage{arydshln}
\usepackage{textcomp}

\newcommand{\bkappa}{\bm{\kappa}}

\newcommand{\bu}{\bm{u}}

\newcommand{\bU}{\bm{U}}

\begin{document}

\title{Observation of topological valley modes in an elastic hexagonal lattice}
\author{Javier Vila}
\email[]{javier.vila@aerospace.gatech.edu}
\affiliation{School of Aerospace Engineering, Georgia Institute of Technology}
\author{Raj Kumar Pal}
\affiliation{School of Aerospace Engineering, Georgia Institute of Technology}
\author{Massimo Ruzzene}
\affiliation{School of Aerospace Engineering and School of Mechanical Engineering, Georgia Institute of Technology}

\date{\today}

\begin{abstract}
We report on the experimental observation of topologically protected edge waves in a two-dimensional elastic hexagonal lattice. The lattice is designed to feature K point Dirac cones that are well separated from the other numerous elastic wave modes characterizing this continuous structure. We exploit the arrangement of localized masses at the nodes to break mirror symmetry at the unit cell level, which opens a frequency bandgap. This produces a non-trivial band structure that supports topologically protected edge states along the interface between two realizations of the lattice obtained through mirror symmetry. Detailed numerical models support the investigations of the occurrence of the edge states, while their existence is verified through full-field experimental measurements. The test results show the confinement of the topologically protected edge states along pre-defined interfaces and illustrate the lack of significant backscattering at sharp corners. Experiments conducted on a trivial waveguide in an otherwise uniformly periodic lattice reveal the inability of a perturbation to propagate and its sensitivity to backscattering, which suggests the superior waveguiding performance of the class of non-trivial interfaces investigated herein.
\end{abstract}


\maketitle

\section{Introduction}

Wave propagation in periodic media has been an active field of research for the past few decades. Energy transport by waves arise in multiple areas of physics as it relates to acoustic, elastic, electromagnetic, electronic and opto-mechanical media. Unique phenomena like negative refraction, directional propagation, focusing and cloaking have been pursued through careful engineering of the band structure, which is a unifying theme for exploration in this diverse set of physical domains. Recently, the advent of topological mechanics~\cite{huber2016topological} has provided an effective framework for the pursuit of robust wave propagation which is protected against perturbations and defects. 
Topologically protected edge wave propagation was originally envisioned in quantum systems and has 
quickly evolved to other classical areas of physics, such as acoustics~\cite{brendel2017snowflake}, photonics~\cite{khanikaev2013photonic}, mechanics~\cite{mousavi2015topologically,Raj2017b} and
optomechanics~\cite{peano2014topological}. In all of these different media, properties such as lossless propagation, existence of waves confined to a boundary or interface, immunity to backscattering and localization in the presence of defects and imperfections are the result of band topology. This makes them classical analogues of topological insulators that support the propagation of topologically protected edge waves (TPEWs).

There are two broad ways to realize topologically protected wave propagation in elastic media. The first one uses active components, thereby mimicking the quantum Hall effect. Changing of the parity of active devices or modulating of the physical properties in time, for example, has shown to alter the direction and nature of edge waves~\cite{swinteck2015bulk,nassar2017modulated}. Examples include magnetic fields in biological systems~\cite{prodan2009topological}, rotating disks~\cite{nash2015topological} and acoustic circulators operating on the basis of a flow-induced bias~\cite{khanikaev2015topologically}. The second way uses solely passive 
components and relies on establishing analogues of the quantum spin Hall effect. These media feature both forward and backward propagating
edge modes, which can be induced by an external excitation of appropriate polarization. The concept is illustrated in several studies by way of both 
numerical~\cite{mousavi2015topologically,pal2016helical,he2016topological} and 
experimental~\cite{susstrunk2015observation,ningyuan2015time} investigations, which involve coupled pendulums~\cite{susstrunk2015observation}, 
plates with two scale holes~\cite{mousavi2015topologically} and resonators~\cite{pal2016helical}, 
as well as electric circuits~\cite{ningyuan2015time}.  Numerous studies have also been conducted on localized non-propagating 
deformation modes at the interface of two structural lattices~\cite{prodan2017dynamical,Raj2017b,chaunsali2017demonstrating}. 
These modes depend on the topological properties of the bands, which in $1D$ lattices are characterized by topological invariant called Zak phase~\cite{xiao2015geometric}. In $2D$ and $3D$ lattices, several researchers have investigated the presence of floppy modes of motion due to nontrivial 
topological polarization and exploited these modes to achieve localized buckling and directional response~\cite{kane2014topological,paulose2015selective,rocklin2016mechanical,rocklin2016directional}.  
In spite of the intense level of activity in this area, to the best of our knowledge, studies reporting on the experimental observation of TPEWs in continuous elastic media have so far been limited. Unique challenges in elastic systems exist due to their high modal densities, which complicate the analysis and design of the band structure and the effective achievement of non-trivial topologies. These also often lead to complex arrangements of materials and intricate connectivities that may be hard to realize in practice. A promising avenue in this regard is the use of valley degrees of freedom as originally envisioned in quantum systems like graphene bilayers~\cite{xiao2007valley,zhang2011spontaneous,zhang2013valley,lu2016observation}. The concept has also been adopted in classical areas such as photonics~\cite{ma2016all}, acoustics and phononics~\cite{Brendel2017,Raj2017b} .

The objective of this study is to exploit valley degrees of freedom to obtain and demonstrate experimentally TPEWs in continuous elastic media. The considered configuration consists of an elastic hexagonal lattice on which concentrated masses are attached at the sub-lattice sites. This provides a simple assembly that is characterized by the symmetry conditions sufficient to open a topologically non-trivial bandgap. The addition of masses at selected locations within a unit cell breaks the $C_{3v}$ symmetry inherent to the hexagonal geometry, while preserving the $C_3$ symmetry. Exploiting the arrangement of masses conveniently leads to lattices that exhibit different topological properties of the bands. When two such lattices wih different topological properties are joined together, TPEWs propagate along the shared interface.

The outline of this paper is as follows: Sec.~\ref{Sec.Theory} explains the concept of valley modes, while Sec.~\ref{Sec. Hex Theory} presents the description of the continuous hexagonal lattice, along with its dispersion analysis. Section~\ref{Sec: Experiments} describes the experimental setup, the estimation of the dispersion diagrams for the lattices, and results showing TPEWs for two different interfaces. Finally, Sec.~\ref{Sec.Concl} summarizes the conclusions and presents potential future research directions of investigation.

\section{Hexagonal Spring-Mass Lattices and Valley Hall Effect Analogy }\label{Sec.Theory}

\begin{figure}[hbtp]
	\centering
\subfigure[]{\includegraphics[width=0.45\textwidth]{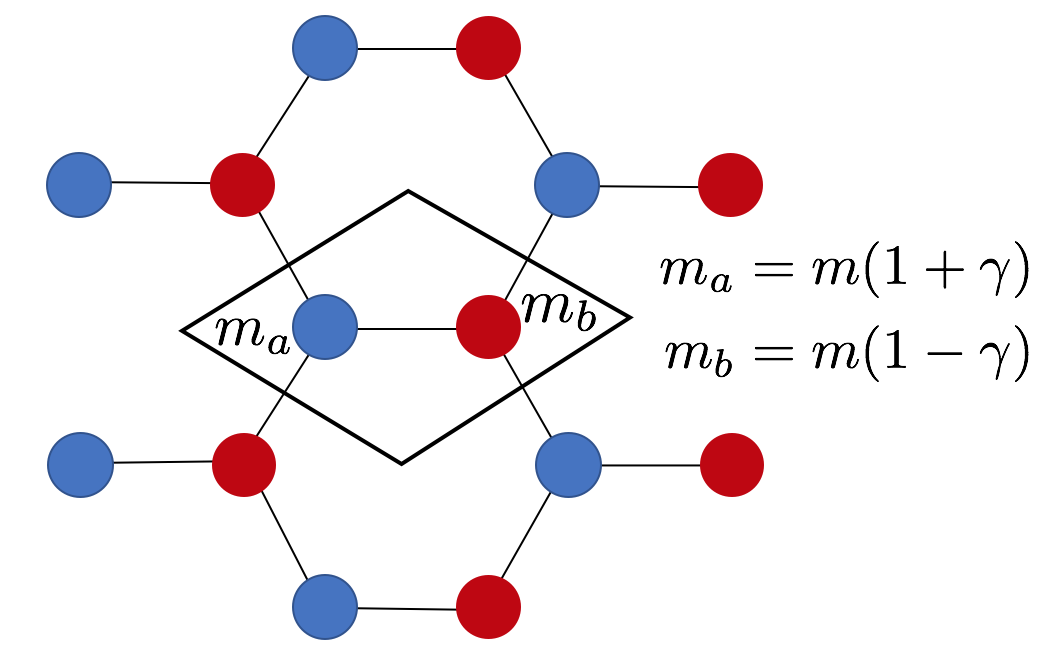}}
\subfigure[]{\includegraphics[width=0.45\textwidth]{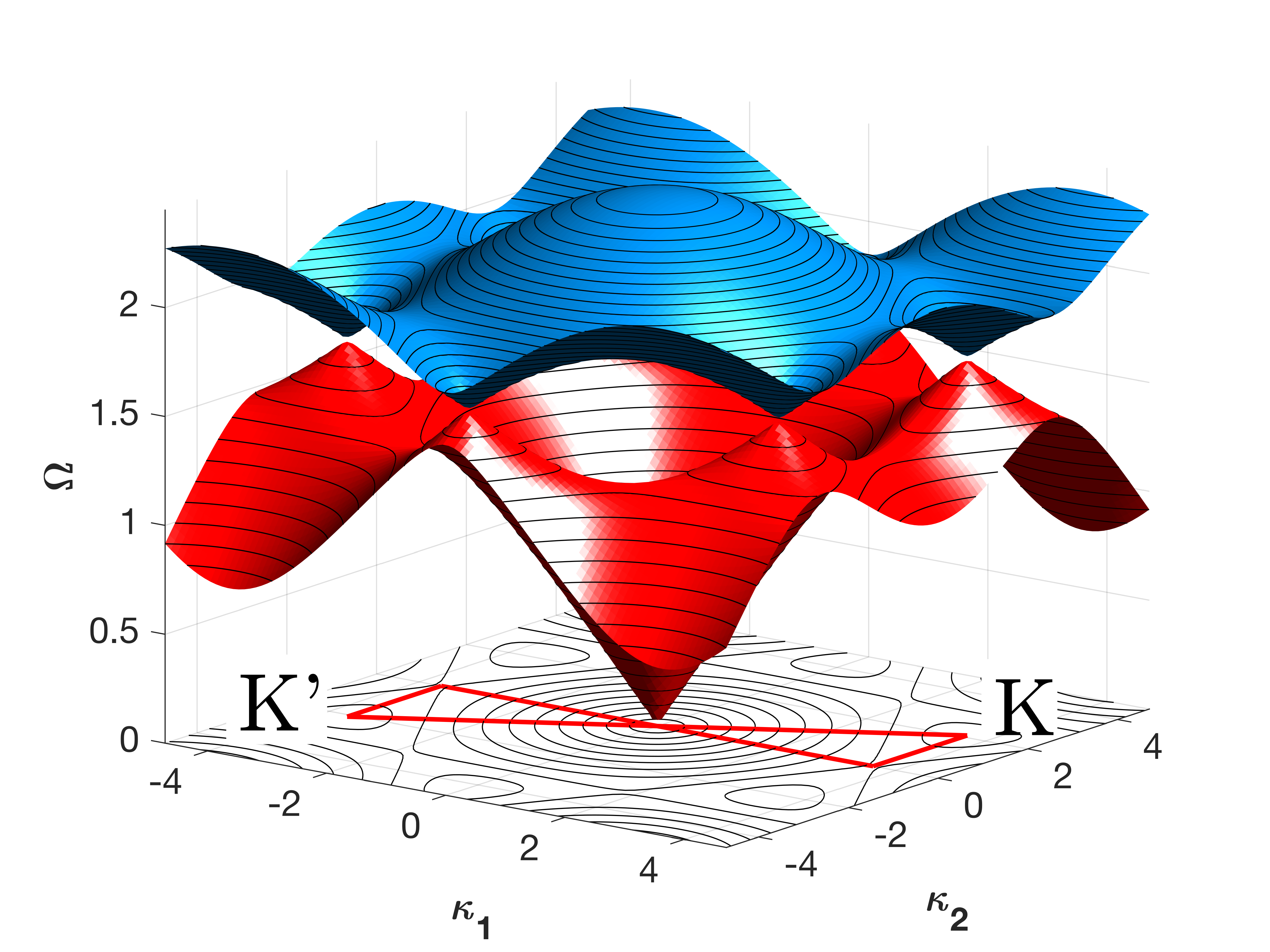}}\\
\subfigure[]{\includegraphics[width=0.45\textwidth]{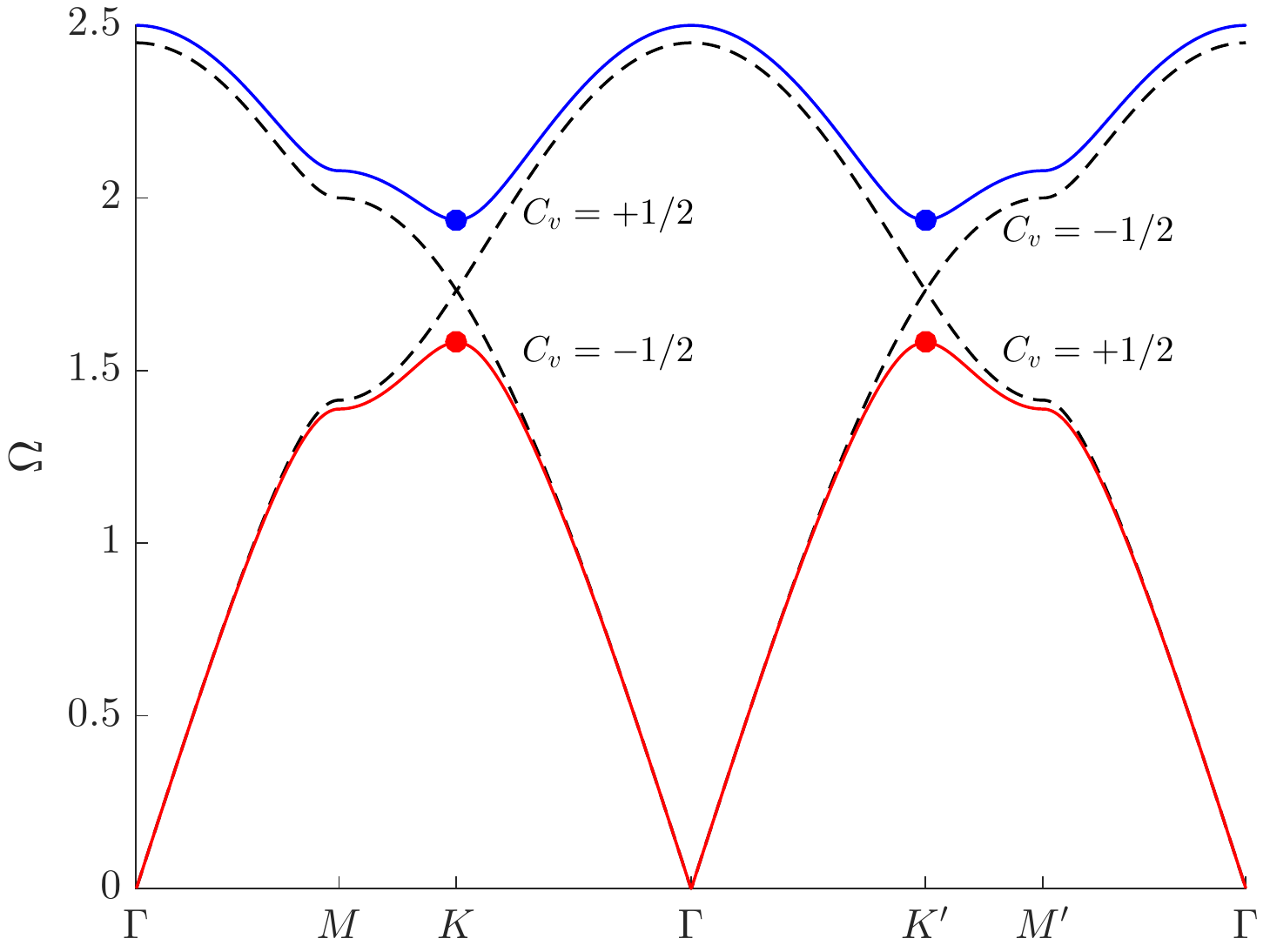}}
\subfigure[]{\includegraphics[width=0.45\textwidth]{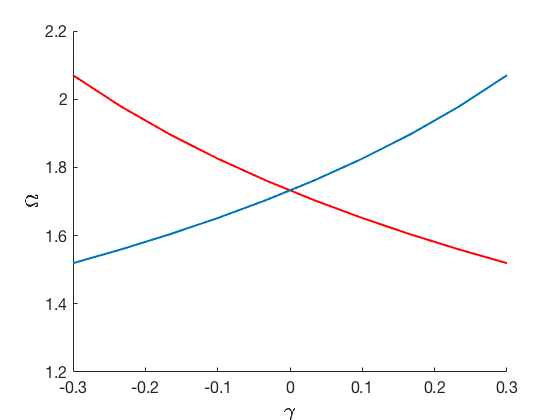}}
 \caption{Hexagonal discrete lattice emulating the QVHE (a). Dispersion surfaces and contours of the first surface highlighting the presence of Dirac cones, the first irreducible Brillouin zone (solid red line), and the location of the K and K$'$ points (b). Band diagrams reporting the values of the valley Chern numbers at the K and K$'$ points (c). Inversion of the bounding frequencies of the bandgap as a function of the parameter $\gamma$ (d).}
	\label{Fig.0}
\end{figure}

We briefly illustrate the Quantum Valley Hall Effect (QVHE) analogy for discrete hexagonal lattices, which provide the basic configuration for the design and subsequent study of the continuous lattices investigated in this paper. While detailed descriptions of the concept can be found in~\cite{Raj2017b}, we summarize here how changes in topological properties can be achieved by considering unit cells that are inverted copies of each other, leading to band inversion in the dispersion diagram, and opposite topological properties at the valley points K and K$'$. In this context, two sufficient conditions guarantee the existence of TPEWs in a periodic system: the unit cell should satisfy $C_3$ symmetry and violate mirror symmetry \cite{xiao2007valley,ma2016all,Chen2016,lu2016}. For example, the discrete hexagonal lattice of Fig.~\ref{Fig.0}a, comprises point masses at the sub-lattice sites connected by linear springs of stiffness $k$ joining nearest neighbors. Each unit cell has two sites $a,b$ where the masses respectively are $m_a = m (1+\gamma) $ and $m_b= m (1-\gamma)$. Thus, the lattice satisfies $C_3$ symmetry but violates mirror symmetry about the lattice vectors $C_{3v}$. Two lattice types can be conveniently obtained by considering values of $\gamma>0$ or $\gamma<0$, which corresponds to switching the position of the masses through a mirror symmetry operation. The band structure of the lattice with equal masses ($\gamma=0$)  shown in Fig.~\ref{Fig.0}b reveals the presence of Dirac cones at the high symmetry points. The symmetry is also highlighted by the iso-frequency contours of the first dispersion surface, on which the path along the boundaries of the first irreducible Brillouin zone and its symmetric counterpart are represented (red solid line), along with the location of the K and K$'$ points. Dimensionless frequency $\Omega=\omega/\omega_0$ is obtained for every wavevector $(\kappa_1,\kappa_2)$, where $\omega_0=\sqrt{k/m}$. The band diagrams obtained for the wave vector tracing these boundaries are shown in Fig.~\ref{Fig.0}c, where comparisons are obtained for the symmetric case with equal masses ($\gamma=0$, dashed curves), and when mirror symmetry is broken ($\gamma \neq 0$, in this case $\gamma=+0.2$, solid curves), which results in the opening of the Dirac cones to form a bandgap. The bounding frequencies of the gap vary as a function of $\gamma$, and a band inversion occurs when $\gamma$ changes in sign, which corresponds to the case of two mirror-symmetric unit cells (see Fig.~\ref{Fig.0}d).

Expressing the Hamiltonian of this lattice in the basis of an extended vector, which is obtained by combining the eigenvectors at the K and K$'$ valley points, illustrates the analogy with quantum valley Hall effect~\cite{Raj2017b,Brendel2017}. Alternatively, the topological properties of the vector bundle associated with the eigenvectors $\bu_0(\bkappa)$ can be used to infer the presence of interface modes \cite{Carpentier2014}. The valley Chern number, which is the integral of the Berry curvature over half the Brillouin zone, characterizes the topology of this vector bundle, see~\cite{xiao2007valley,zhang2013valley}. In hexagonal lattices with broken mirror symmetry, the valley Chern number takes values $\pm1/2$ at the opposite valleys, i.e. the K and K$'$ points in reciprocal space, indicating opposite polarization of the corresponding eigenmodes~\cite{Chen2016} (Fig.~\ref{Fig.0}c). In the described discrete lattice (see Fig. \ref{Fig.0}a) the valley Chern number of the first mode is $(-)1/2$ at $($K$)$K$'$ points for $\gamma>0$ and vice-versa for $\gamma<0$~\cite{Raj2017b}. To realize TPEWs, it suffices to build a structure in which two lattices with opposite valley Chern numbers share an interface. These two lattices may have the same band structure but their eigenmodes at the valley points have opposite polarization. When these two lattices share a common interface, topologically protected localized modes exist at frequencies within the bandgap~\cite{Raj2017b}, and TPEWs can propagate confined to that interface.

\section{Continuous Hexagonal Lattice}\label{Sec. Hex Theory}

\subsection{Configuration and material properties}
\begin{figure}[hbtp]
	\centering
\includegraphics[width=0.4\textwidth]{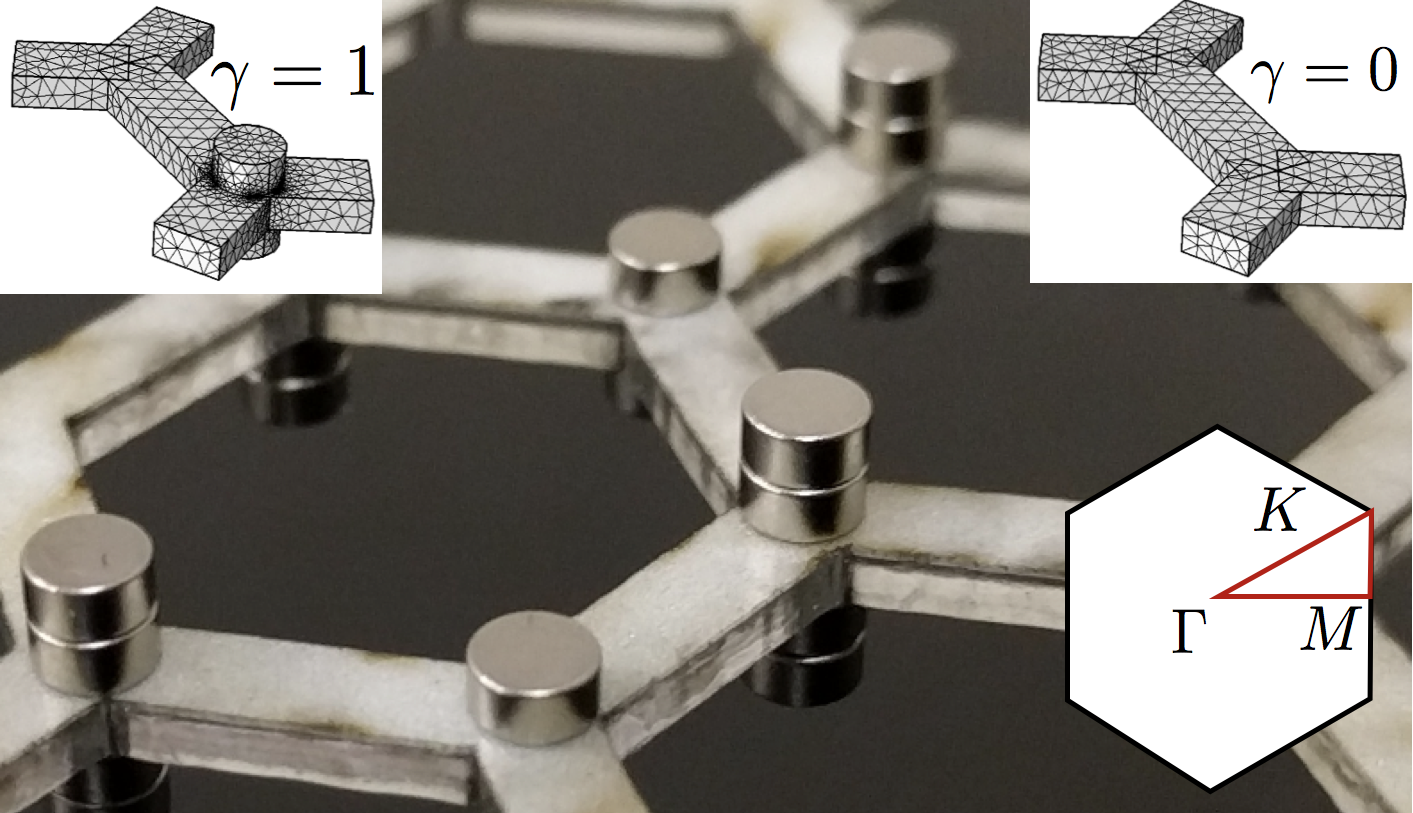}
\caption{Experimental lattice with added masses at the sub-lattice sites. Insets show the FE discretization of the unit cell for numerical study for configurations defined by $\gamma=0$ and $\gamma=1$, as well as the first and irreducible Brillouin zone (red dashed line) for the lattice.}
	\label{Fig.lattice experiment}
\end{figure}

The characteristics of the conceptual lattice summarized above guide the design of the continuous hexagonal elastic lattice of Fig.~\ref{Fig.lattice experiment}. The lattice is fabricated out of a square acrylic panel of side $308.4 \,\text{mm}$ and thickness of $1.59 \,\text{mm}$. The side of the each hexagon measures $L = 10.7 \,\text{mm}$, while  the width of the beams is $w= 3.2 \,\text{mm}$. The masses consist of cylindrical nickel-plated neodymium magnets ($\rho_c=7400 \,\text{kg}/\text{m}^3$, $E_c=41 \,\text{GPa}$ and $\nu_c=0.28$) of height $1.5 \,\text{mm}$ and diameter $3.2 \,\text{mm}$. The material properties of acrylic are: density $\rho=1190 \,\text{kg}/\text{m}^3$, Young's modulus $E=3.2 \,\text{GPa}$, Poisson's ratio $\nu=0.35$. The lattice is generated by a set of lattice vectors $\bm a_1 = \sqrt{3} L \,\, [\sqrt{3}/2, \,\, -1/2]$ and $\bm a_2 = \sqrt{3} L \,\, [\sqrt{3}/2, \,\, 1/2]$.  Unit cell mirror symmetry is broken by adding cylindrical masses at selected sub-lattice sites $a,b$, in analogy with the discrete lattice in Fig.~\ref{Fig.0}a. The mass added by the cylinders is defined respectively as $m_a=\left( |\gamma|+\gamma\right) m_c, m_b=\left( |\gamma|-\gamma\right) m_c$, where $m_c$ denotes the mass of one cylinder and $\gamma$ is the parameter chosen to define both magnitude and location of the added mass at each site.  Of note is the fact that the lattice is a continuous structure that is characterized by inherent mass properties defined by the density of the material. Therefore the terms $m_a, m_b$ denote the added mass at the sub-lattice sites. An even number of cylinders are added at each location in order to preserve symmetry in the thickness direction, and for practical purposes in the experimental implementation of the concept, whereby attracting magnetic cylinders are clamped at the desired location. Hence, values $\gamma>0$ describe the addition of masses at site $a$, while $\gamma<0$ corresponds to an added mass in $b$. In addition, the case $\gamma=+1 (-1)$ describe the addition of two cylinders in $a (b)$, and finally when $\gamma=0$ corresponds to the case of no additional masses.

\subsection{Dispersion Analysis}
\begin{figure}[hbtp]
	\centering
\subfigure[]{\includegraphics[width=0.4\textwidth]{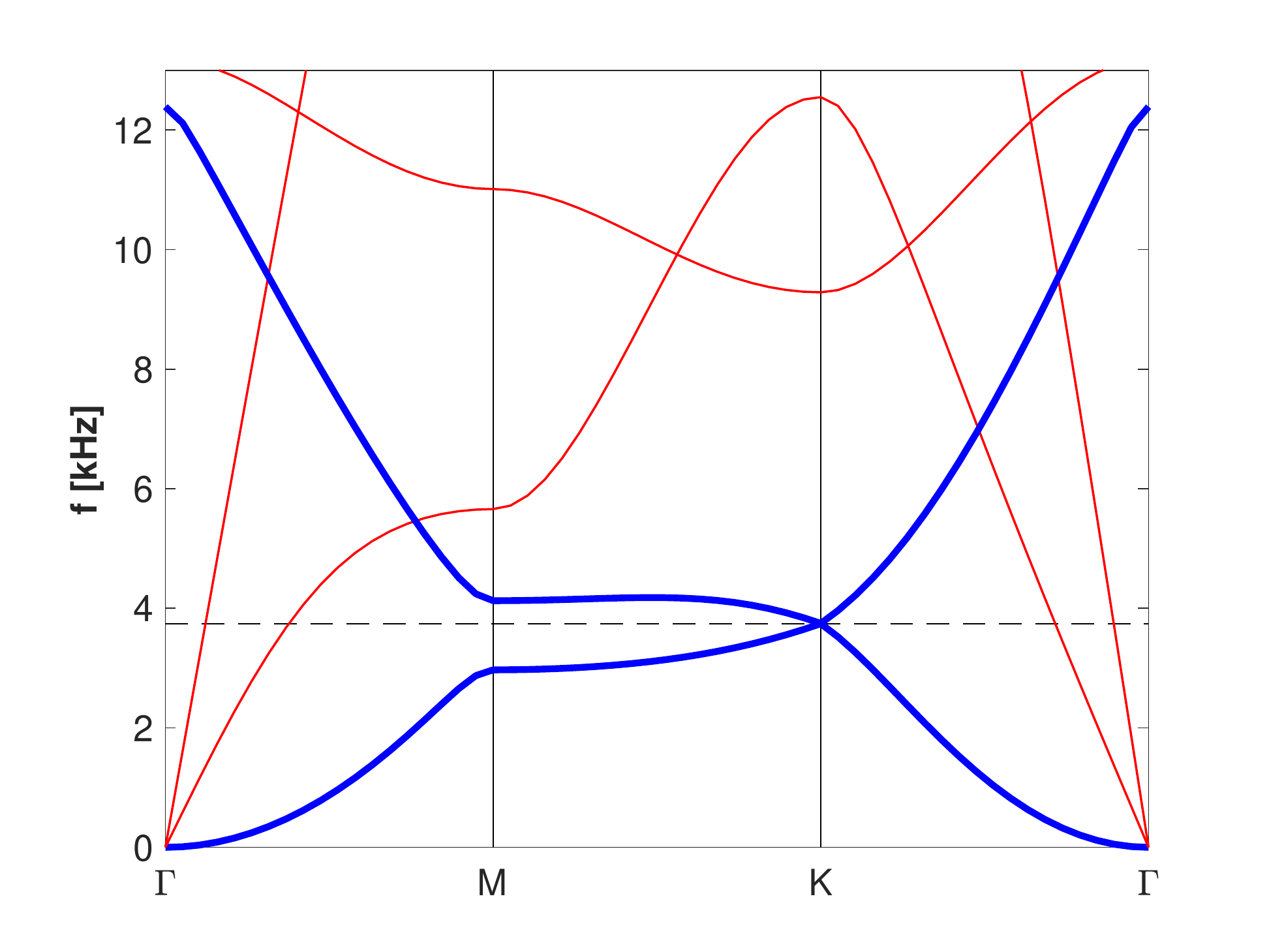}}
\subfigure[]{\includegraphics[width=0.4\textwidth]{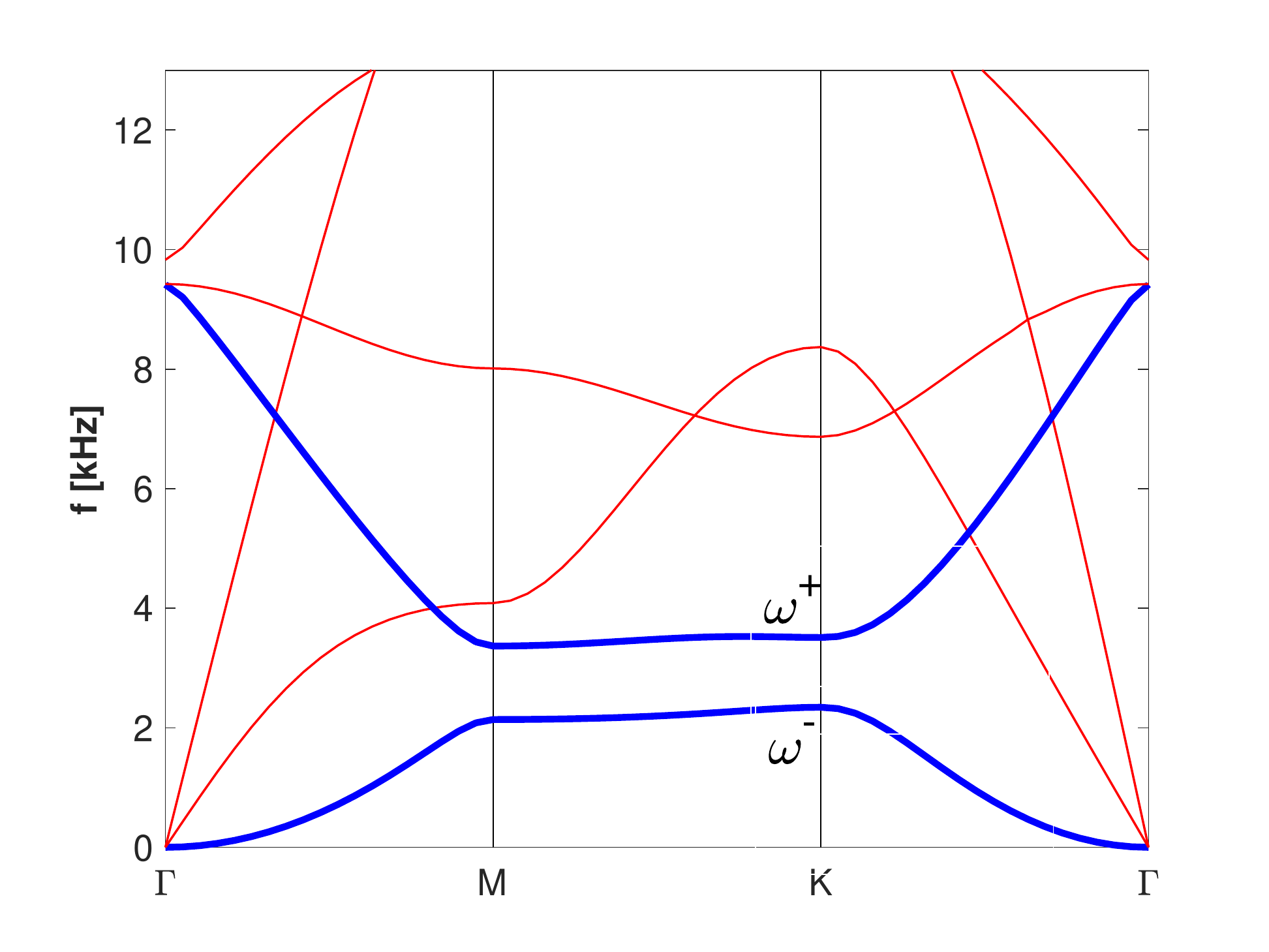}}\\
\subfigure[]{\includegraphics[width=0.4\textwidth]{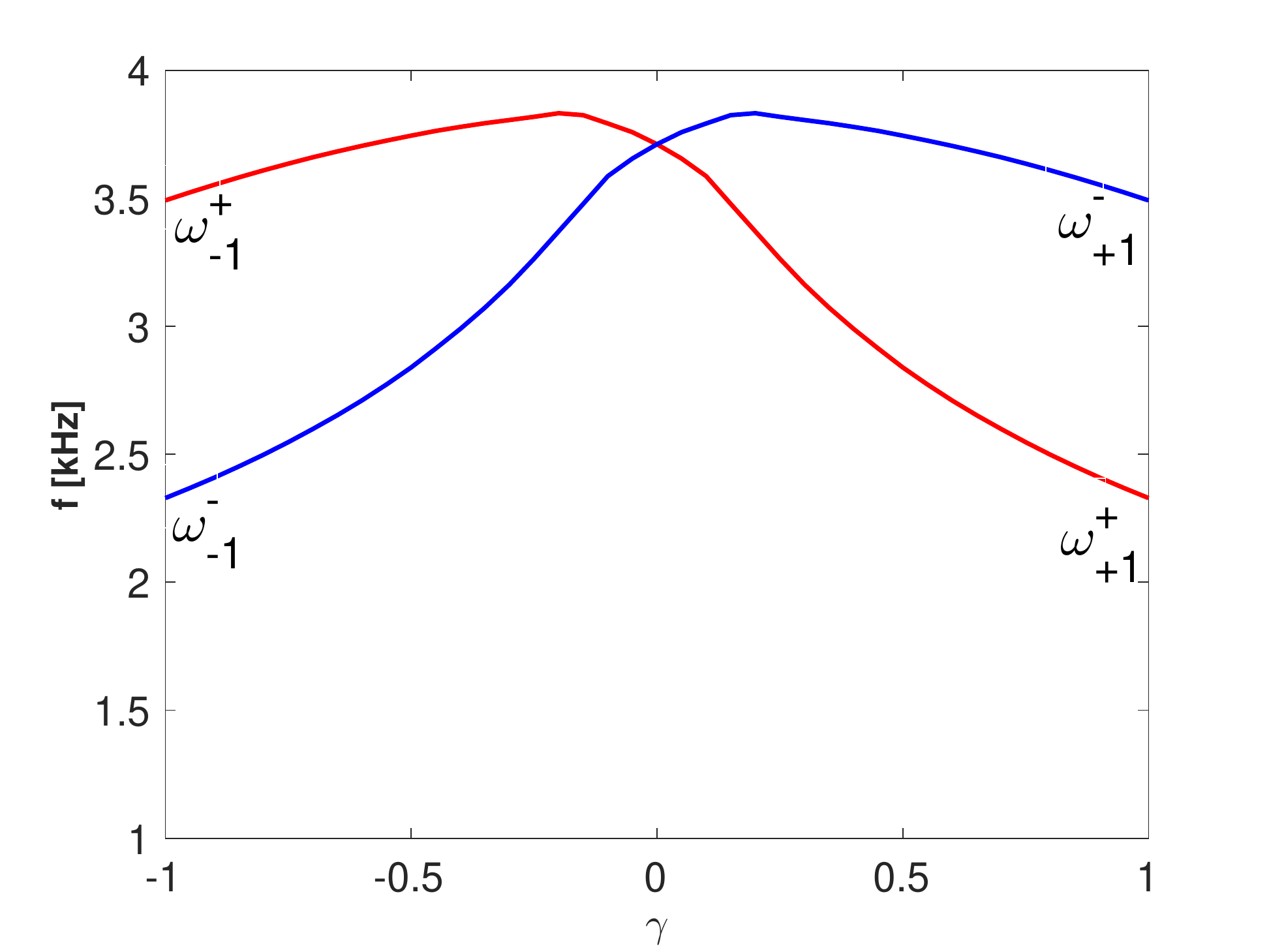}}
\subfigure[]{\includegraphics[width=0.3\textwidth]{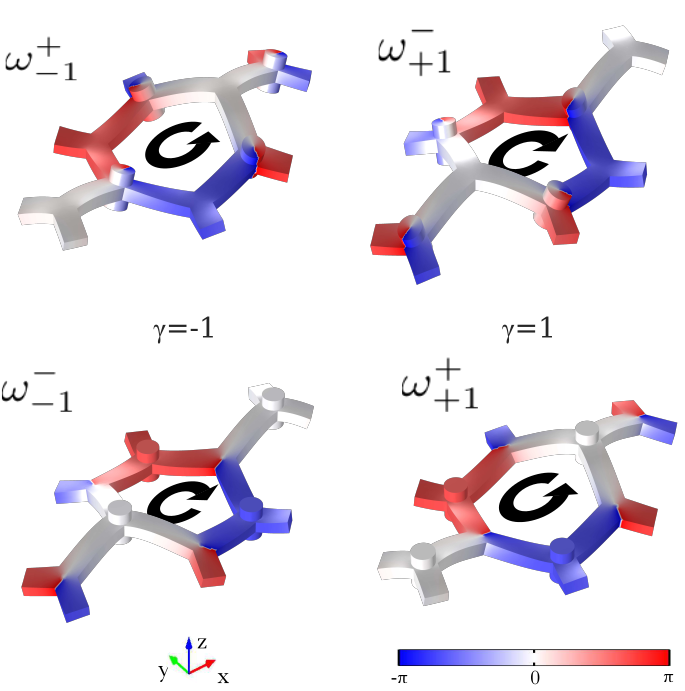}}
	\caption{Dispersion diagrams: lattice without masses, $\gamma=0$ (a); lattice with masses, $\gamma=1$ (b) ((black) thin dashed line: location of Dirac point, (blue) thick lines: out-of-plane wave mode, (red) thin lines: shear polarized modes). Variation of bandgap bounding frequencies $\omega^+$ (red line),  $\omega^-$ (blue line) at K as a function of $\gamma$ showing band inversion (c), and  phase of corresponding eigenfunctions of the first two out-of-plane modes at K for $\gamma=-1$ and $\gamma=1$ (d).}
	\label{Fig.FE_DispDiagrams}
\end{figure}

The dispersion properties of the lattice are estimated based on the Finite Element (FE) discretization of a unit cell performed within the COMSOL Multiphysics environment. Each unit cell is discretized using around 20000 second-order tetrahedral elements, and the considered mesh is illustrated in the insets of Fig.~\ref{Fig.lattice experiment}. The underlying behavior of the lattice is governed by the standard equations of linear elasticity for an isotropic medium \cite{achenbach2012}, 
\begin{equation}\label{Equilibrium}
\rho\ddot\bu-\left[ \left( \lambda+\mu\right)\nabla\left( \nabla\bu\right) +\mu\nabla^2\bu\right]=0,
\end{equation}
where $\bu$ is the displacement and $\lambda$, $\mu$ are the Lam\'{e} constants of the solid. Imposing a plane wave solution and the enforcement of Floquet-Bloch conditions to the unit cell degrees of freedom leads to a linear eigenvalue problem that is solved in terms of frequency for an assigned wave vector. The dispersion properties are evaluated along the edge of the first irreducible Brillouin zone for the hexagonal lattice under consideration. Results for lattices characterized by $\gamma=0$ and $\gamma=1$ are shown in Fig.~\ref{Fig.FE_DispDiagrams}a,b.

The dispersion analysis predicts multiple wave modes that correspond to the numerous degrees of freedom provided by the considered FE mesh. Each node has 3 degrees of freedom, which is reflected by the three branches emanating from the $\Gamma$ point at zero frequency. The study focuses on the wave mode that is dominated by the out-of-plane motion component, which is characterized by a parabolic dispersion branch at long wavelengths, in contrast to the shear polarized modes that in this range are non dispersive. In the diagrams of Fig.~\ref{Fig.FE_DispDiagrams}, the mode of interest is represented by a thick solid (blue) line, while all other modes are represented by the thin (red) lines. Similar to the discrete lattice, the hexagonal lattice with no masses attached ($\gamma=0$) has $C_{3v}$ symmetry: a Dirac point is observed at the K point of the reciprocal lattice space (Fig.~\ref{Fig.FE_DispDiagrams}a) and identified by the horizontal dashed line. In contrast, adding 2 cylindrical masses at the $a$ site ($\gamma=1$) breaks mirror symmetry, and produces a bandgap between the first and the second out-of-plane modes, see Fig. \ref{Fig.FE_DispDiagrams}b. The bounding frequencies of the gap, denoted $\omega^+$ and $\omega^-$ are tracked as a function of $\gamma$, which leads to the plot of Fig. \ref{Fig.FE_DispDiagrams}c showing a band inversion. The eigenfunctions $\bU$ associated to these eigenvalues for $\gamma=-1$ and $\gamma=1$ are depicted in Fig.~\ref{Fig.FE_DispDiagrams}d. We observe that, while the eigenvalues are preserved under the transformation  $\gamma \to -\gamma$ due to time reversal symmetry, the eigenfunctions feature different polarizations that reflect the mirror-symmetry relations of the corresponding unit cells. The transformation $\gamma\to -\gamma$ may be achieved by simply reversing the direction of the lattice basis vectors, so the positions of the sub-lattice sites $a$ and $b$ are switched. Due to the broken $C_{3v}$ mirror symmetry, a reflection changes the eigenfunctions and thereby the bands topology \cite{Brendel2017}. Let us examine in detail the phase of the eigenfunctions $\bU$ at K point. Note that for $\gamma<0$ the eigenfunction of the first mode has clockwise polarization and the eigenfunction of the second one has counter-clockwise polarization, whereas the opposite is observed for $\gamma>0$. Fig. \ref{Fig.FE_DispDiagrams}c shows that the bands are inverted when $\gamma$ changes sign (at $\gamma=0$). Furthermore, the K$'$ points have opposite polarizations to the K points due to time reversal symmetry. The change in polarization across $\gamma=0$ suggests that lattices with $\gamma>0$ and $\gamma<0$ have opposite valley Chern numbers \cite{Fosel2017,Raj2017b}, and that TPEWs are expected to exist along an interface between a lattice with $\gamma>0$ and a lattice with $\gamma<0$ at frequencies within the common bandgap.

\section{Experimental Results}\label{Sec: Experiments}

\subsection{Experimental Set-up}

The numerical simulations described in the previous section guide the design and experimental characterization of the considered hexagonal lattice. The configuration considered for the numerical studies is fabricated by cutting the lattice out of an acrylic panel according to the dimensions described in Section III. Wave motion in the lattice is induced by a PZT disc bonded at the desired locations, and driven by a voltage signal generated by a signal generator upon amplification. Full-field motion of the lattice is recorded through a scanning Laser Doppler vibrometer, which measures the out-of-plane velocity of points belonging to a predefined measurement grid. While the equipment records one point at a time, repeating the excitation to record the response at measurement location and the tracking of the phase between subsequent measurements allow the recording of the full-field wave motion of the lattice. The measurements include 7 points along the side $L$ of the hexagon, so that measurements are conducted over a total of 3670 points over the entire lattice. After recording, the wavefield data are interpolated on a regular rectangular grid that includes 100 points along the horizontal ($x$) and vertical ($y$) extent of the measurement domain. The excitation consists of broadband frequency pulses that cover the frequency range of interest, which is up to $12$ kHz. This is achieved through modulated sinusoidal pulses and their superposition, or half-cycle pulses whose duration defines the frequency bandwidth of the excitation. Experiments explained in section \ref{Sec.Expt} are conducted on unit cell assemblies defined by $\gamma=0$, or $\gamma = 1$. Experiments showed in section \ref{Sect.Waveguide} are conducted on lattices with an interface between two unit cell types ($\gamma=\pm1$).

\subsection{Estimation of the dispersion properties}\label{Sec.Expt}

The measurement and their subsequent interpolation produce a data set in the form of a matrix $w(x,y,t)$ that describes the evolution of the deflection of the lattice in time. The matrix of the experimental results is analyzed in Fourier space by performing a three-dimensional Fourier transformation (3D-FT), which gives~\cite{Michaels2011}:
\[ \hat{w} (\kappa_x, \kappa_y, \omega) = \mathcal{F}_{3D}[w(x,y,t)] \]
 
The resulting quantity describes the spectral content, both in terms of frequency as well as reciprocal space, of the recorded wavefield. Cross sections along defined wave paths $\mathcal{C}$, i.e. $ \hat{w} (\kappa|_{\mathcal{C}}, \omega)$, illustrate the spectral content as a function of frequency for the wave vector varying along specific lattice directions, while evaluation at one frequency $\omega_0$,  i.e. $\hat{w} (\kappa_x, \kappa_y, \omega_0)$, illustrates the distribution of energy in the reciprocal wavenumber space at that frequency. These maps provide direct visualization of the dispersion characteristics of the domain of interest, and are therefore used to validate the numerical predictions for the lattice under consideration.

\begin{figure}[hbtp]
	\centering
\subfigure[]{\includegraphics[width=0.45\textwidth]{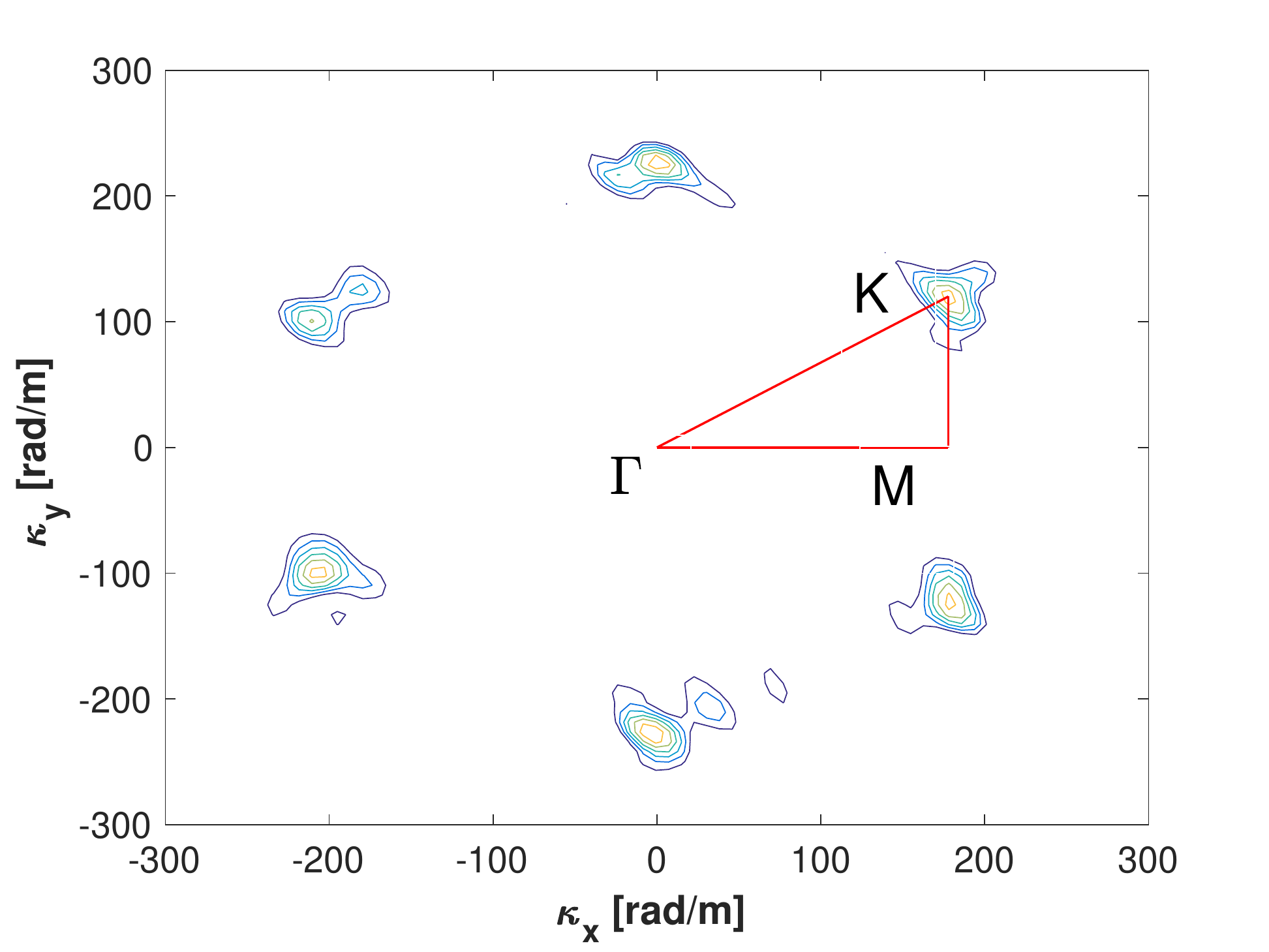}}\\
\subfigure[]{\includegraphics[width=0.45\textwidth]{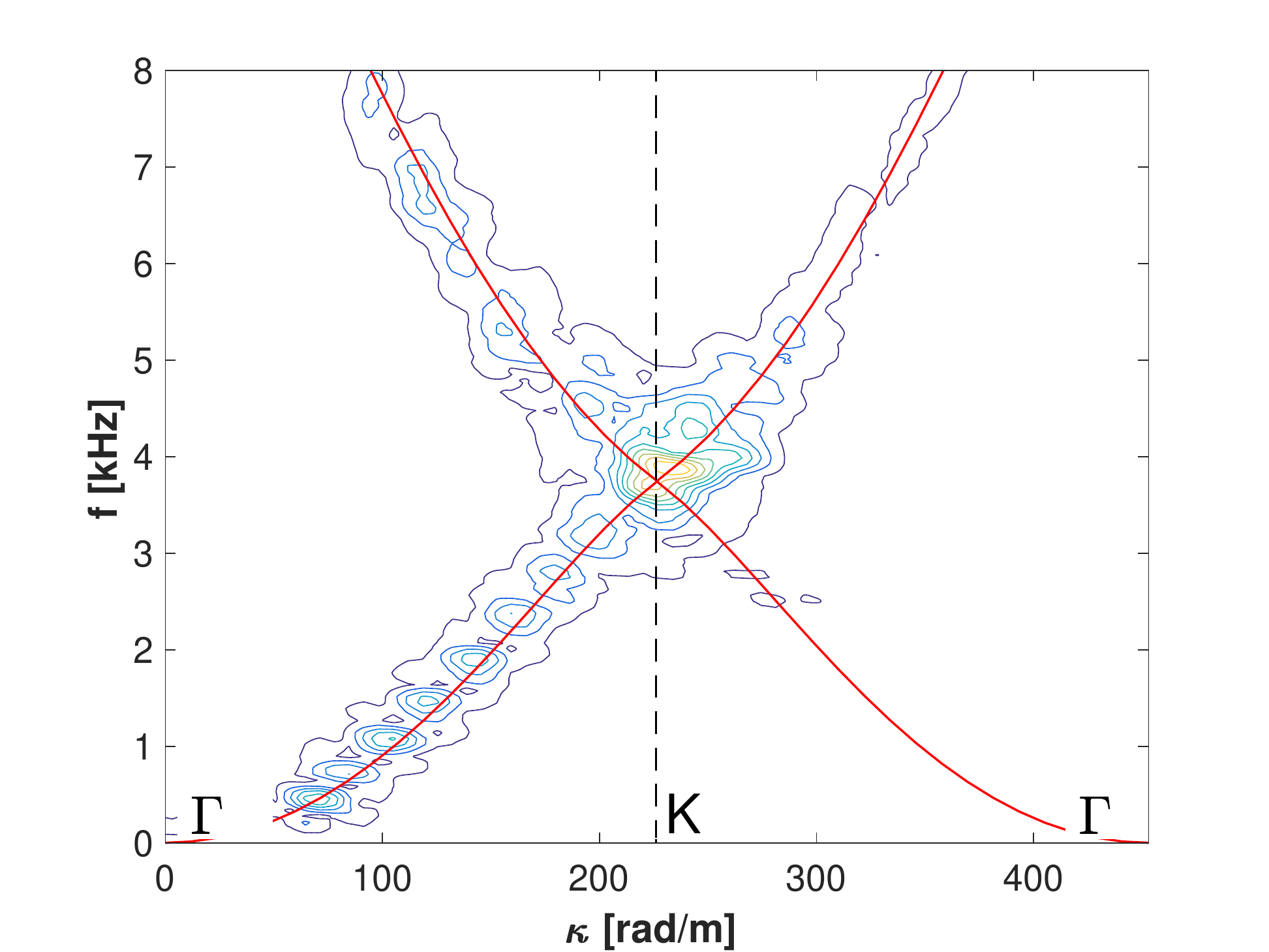}}
\subfigure[]{\includegraphics[width=0.45\textwidth]{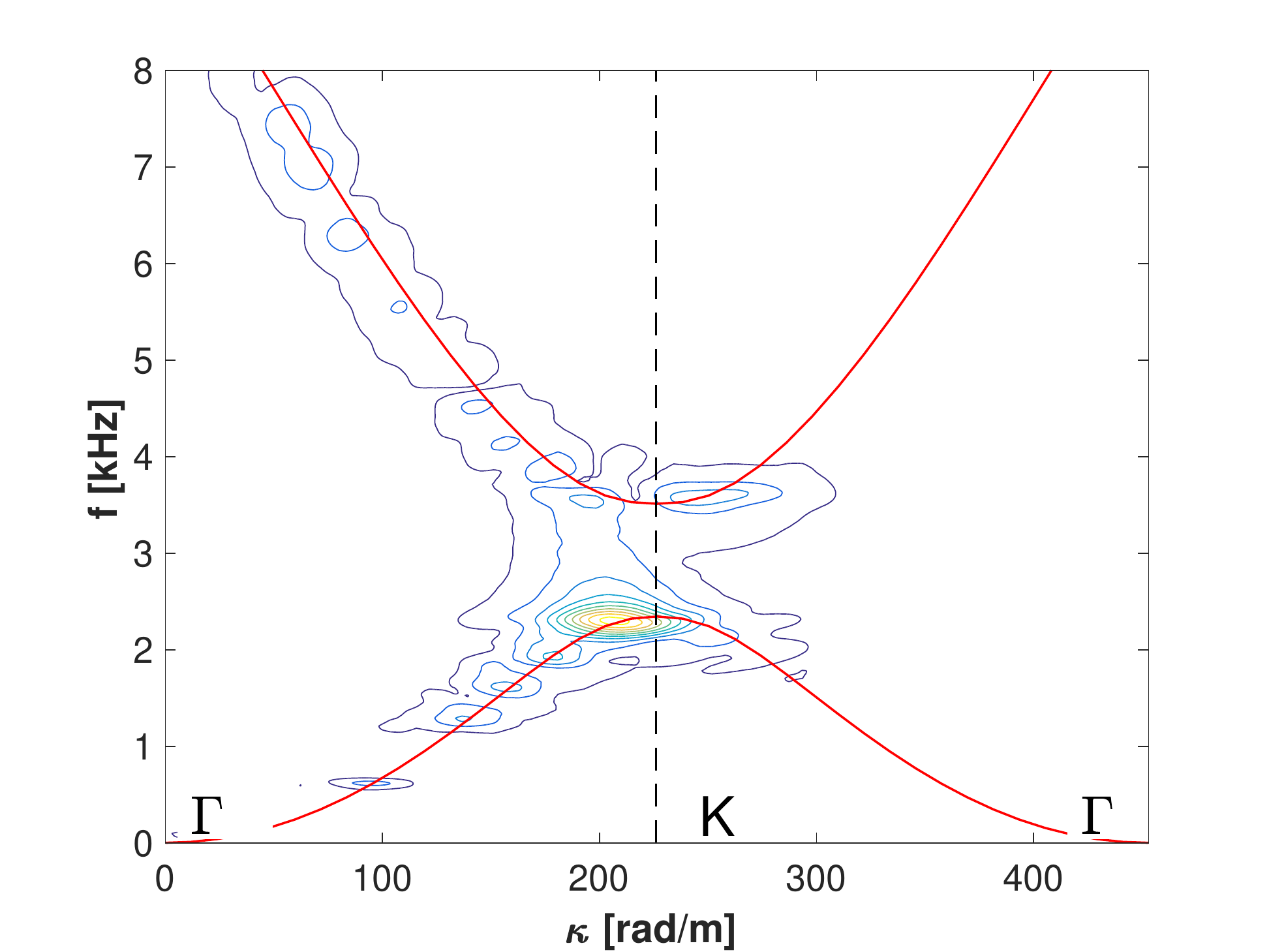}}
	\caption{Experimental 3D-FTs. Lattice with $\gamma=0$: cross section $|\hat{w} (\kappa_x, \kappa_y, \omega_0)|$ at frequency $f_0 \approx 3.75$ kHz close to the numerically predicted Dirac cone (a). Frequency/wavenumber representation $ |\hat{w} (\kappa|_{\mathcal{C}}, \omega)|$ along the path $\mathcal{C}:  \Gamma - K - \Gamma$ and comparison with COMSOL predictions (solid red line): lattice with $\gamma=0$ (b), and $\gamma=1$ (c).}
	\label{Fig.Exp_DispDiagrams}
\end{figure}

We first verify experimentally the dispersion diagrams of the acrylic hexagonal lattice with $\gamma=0$ configuration (no masses attached). The elastic lattice is excited at its center using a piezoelectric patch, which applies a pulse of $40\mu s$ to excite frequencies up to $12.5\,$kHz. Figure~\ref{Fig.Exp_DispDiagrams}a displays the magnitude $|\hat{w} (\kappa_x, \kappa_y, \omega_0)|$ of the 3D-FT at a frequency of $f_0=\omega_0/(2\pi)\approx 3.75\,$ kHz, which is close to the Dirac cone frequency identified by the numerical study of dispersion (see Fig. \ref{Fig.FE_DispDiagrams}a). The contours correspond to the magnitude $|\hat{w}|$, which is not of particular interest here. Most relevant is their location: they are localized at the high symmetry points and effectively illustrate a condition that defines a Dirac point. For reference, the boundaries of the first irreducible Brillouin zone are shown along with the points defining its boundary. The size of the zone is defined by a lattice vector $a =\sqrt{3} L \approx 18.4$ mm, which corresponds to the magnitude of the wave vector at the K point of $\kappa \approx 226$ rad/m. Next, results are presented in terms of frequency/wavenumber content by considering a cross section of the 3D-FT along the path $\mathcal{C}:  \Gamma - K - \Gamma$ for $\gamma = 0$ (no masses added). Figure~\ref{Fig.Exp_DispDiagrams}b illustrates the dispersion branches detected during the experiments, which compare very well with the COMSOL predictions (red solid line). The case of $\gamma=1$ is then tested and Fig.~\ref{Fig.Exp_DispDiagrams}c displays the results in the frequency/wavenumber domain. An opening of the bandgap at the K point is observed as predicted by the COMSOL simulations, again represented by the solid red line superimposed to the contours.

\subsection{Experimental Observation of Topologically Protected Interface Waves}\label{Sect.Waveguide}

The dispersion studies and the experimental set-up developed allow the investigation of the existence of topologically protected modes at the interface of lattices consisting of unit cells that are inverted copies of each other ($\gamma=-1$ and $\gamma=1$). This is easily achieved by placing magnetic cylinders as added masses at the selected locations, so that a variety of interfaces can be introduced and tested.

\begin{figure}[hbtp]
	\centering
\subfigure[]{\includegraphics[width=0.3\textwidth]{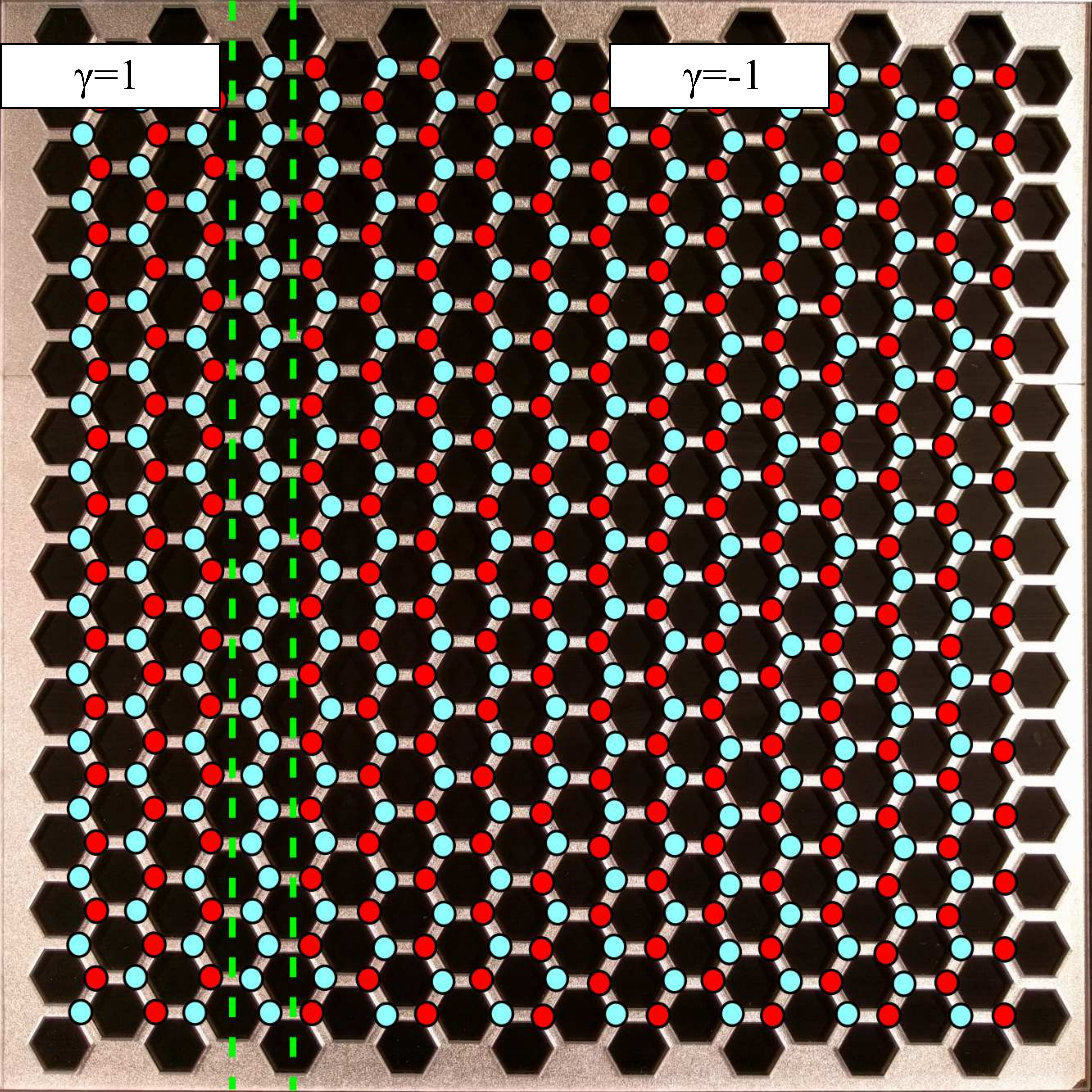}}\\
\subfigure[]{\includegraphics[width=0.3\textwidth]{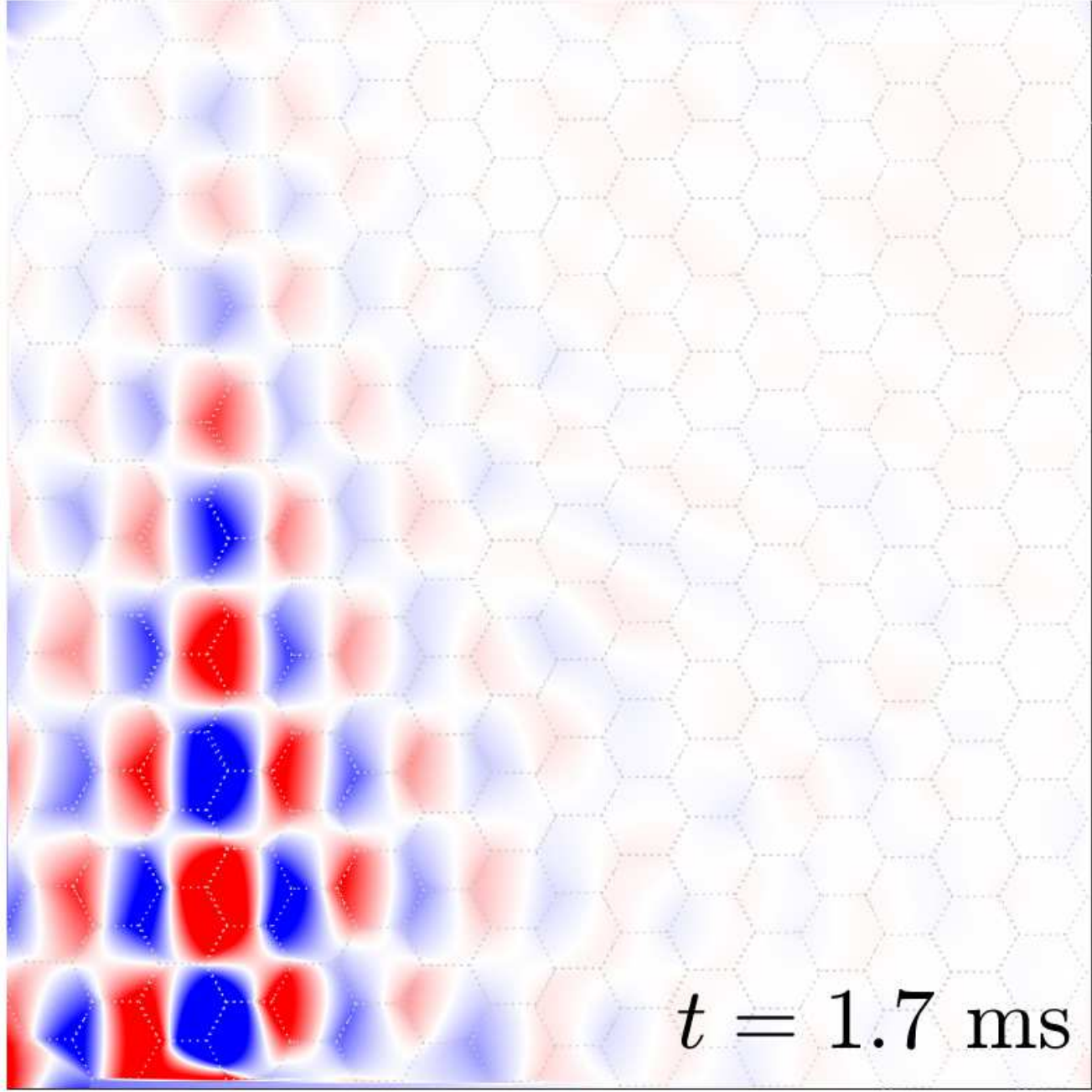}}
\subfigure[]{\includegraphics[width=0.3\textwidth]{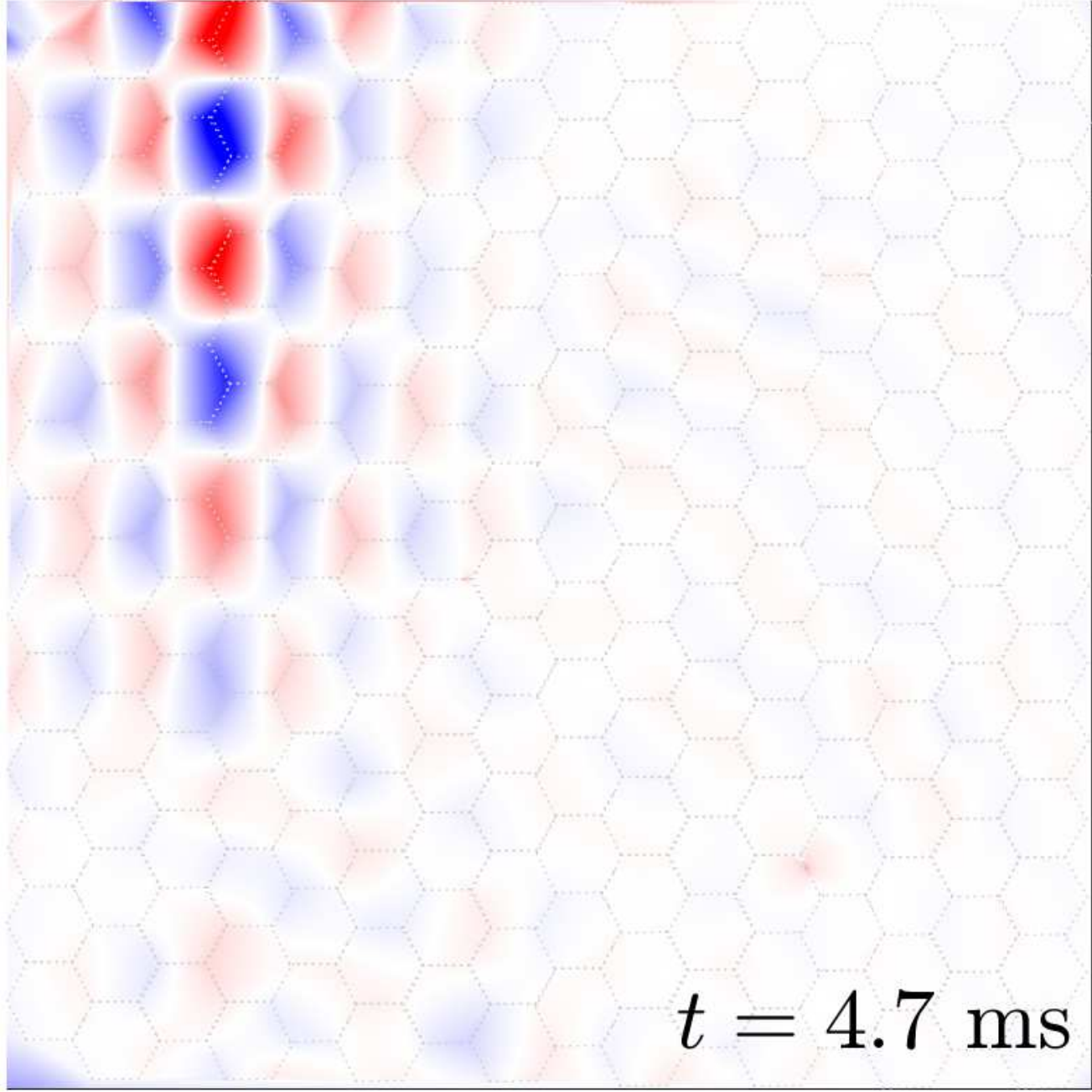}}
\caption{Line interface (red dots indicate two masses attached at the sub-lattice sites and cyan dots denote locations where no masses are added) with $\gamma=(-)1$ on the left(right) (a). Snapshots of measured wave motion at two instants of time. Excitation is a 11 cycles tone burst at $3$ kHz (b).}
	\label{Fig.Waveguide1}
\end{figure}

We first investigate the straight line interface shown in Fig. \ref{Fig.Waveguide1}a. The cylindrical masses are placed so that the unit cells to the left of the interface have $\gamma=1$ and those to the right have $\gamma=-1$. The structure is excited at the intersection between the lower edge of the plate and the interface with a tone burst signal of 11 cycles at a frequency of $3\,$kHz. Figures \ref{Fig.Waveguide1}b and \ref{Fig.Waveguide1}c display time snapshots of the measured out-of-plane displacement by plotting the contours of the interpolated wavefield. The contours are normalized by the maximum displacement amplitude $3\times10^{-8}$ m. Movie files showing time evolution of the displacement field based on actual measurement points are provided en the supplementary material. The figures illustrate how the excitation travels along the interface, and has limited penetration into the bulk. Of interest is the fact that the interface mode is still observable after it reaches the boundary opposite to the excitation location, although the amplitude is reduced by the material dissipation (viscoelastic damping) that is particularly noticeable in the acrylic substrate utilized for the tests. 

\begin{figure}[hbtp]
	\centering
\subfigure[]{\includegraphics[width=0.3\textwidth]{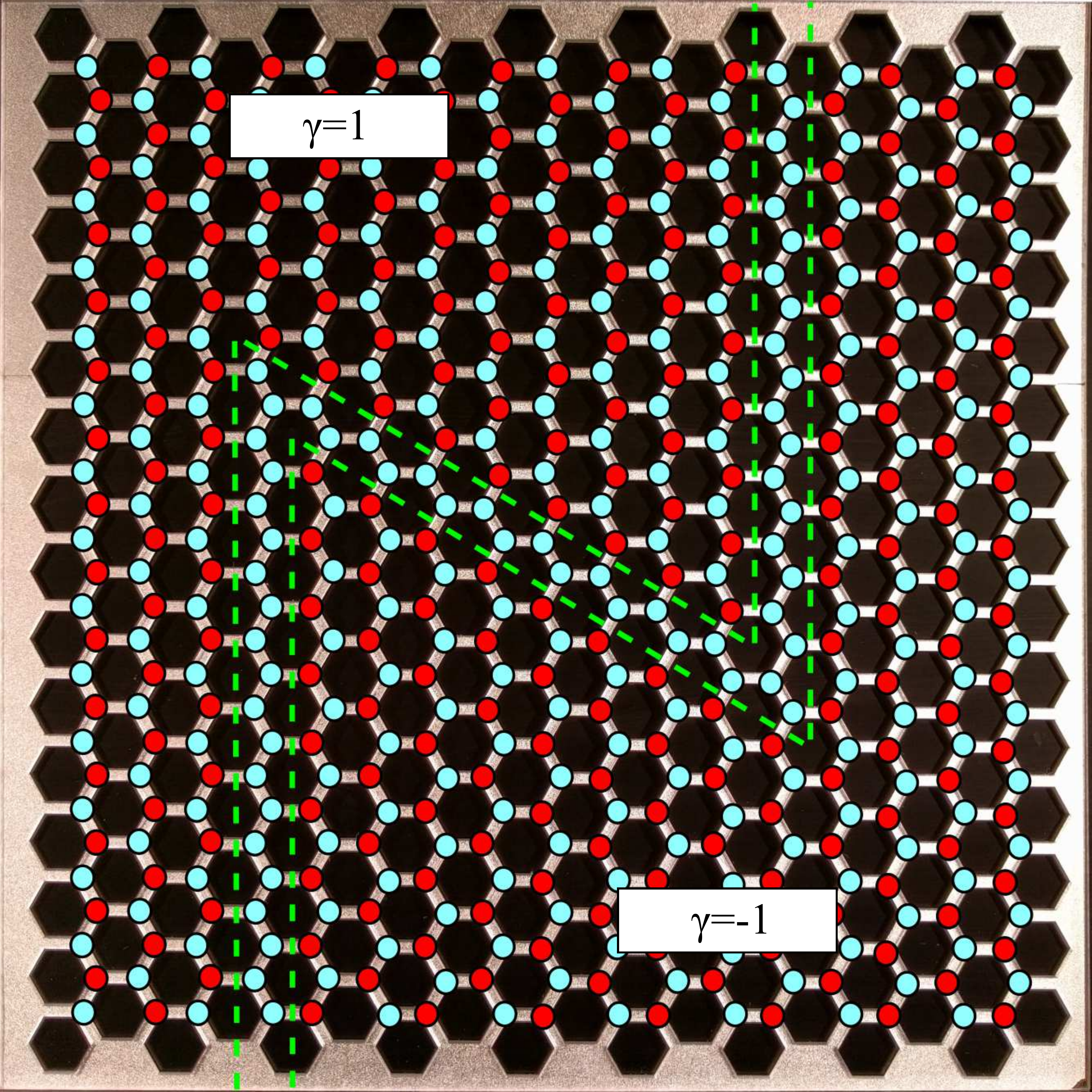}}\\
\subfigure[]{\includegraphics[width=0.3\textwidth]{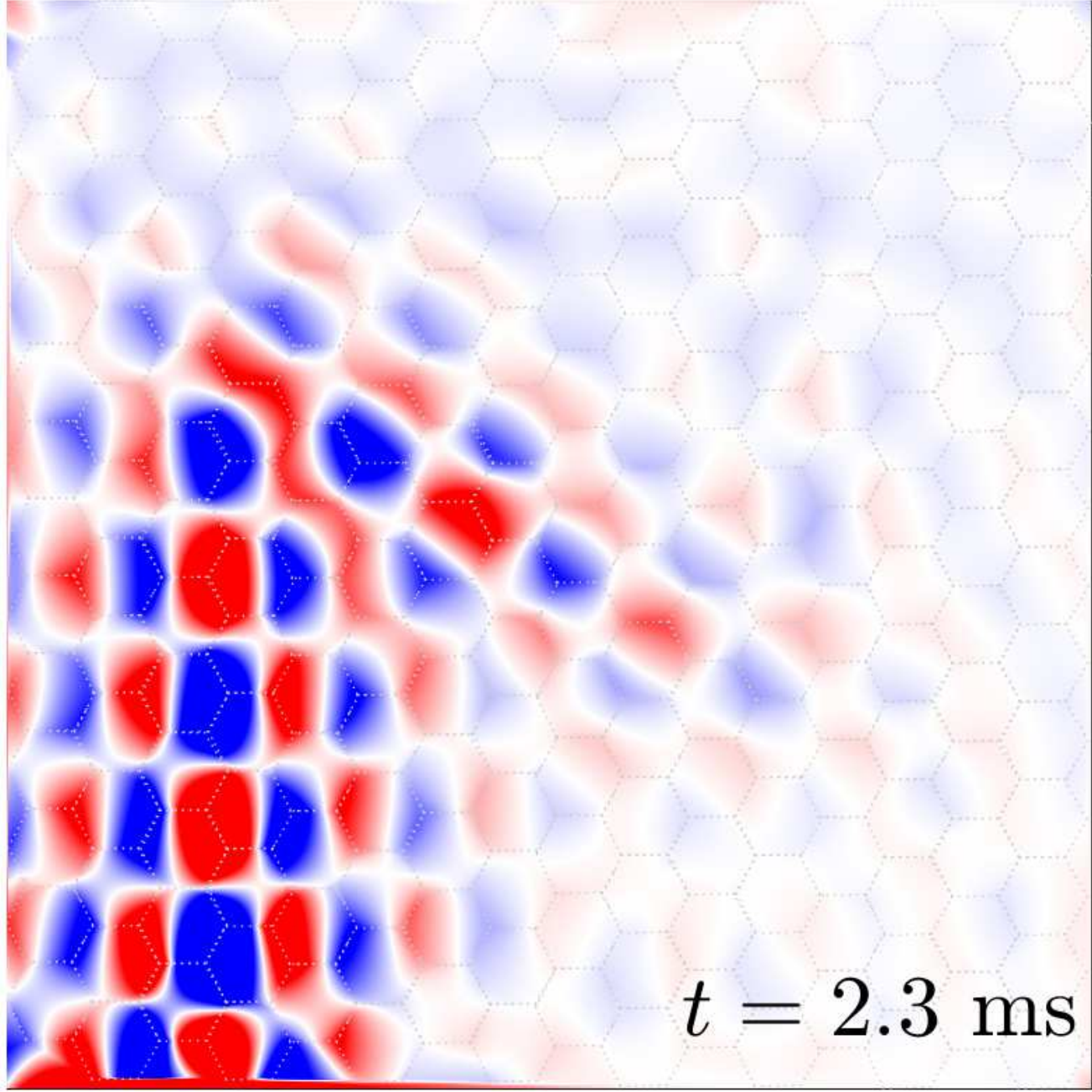}}
\subfigure[]{\includegraphics[width=0.3\textwidth]{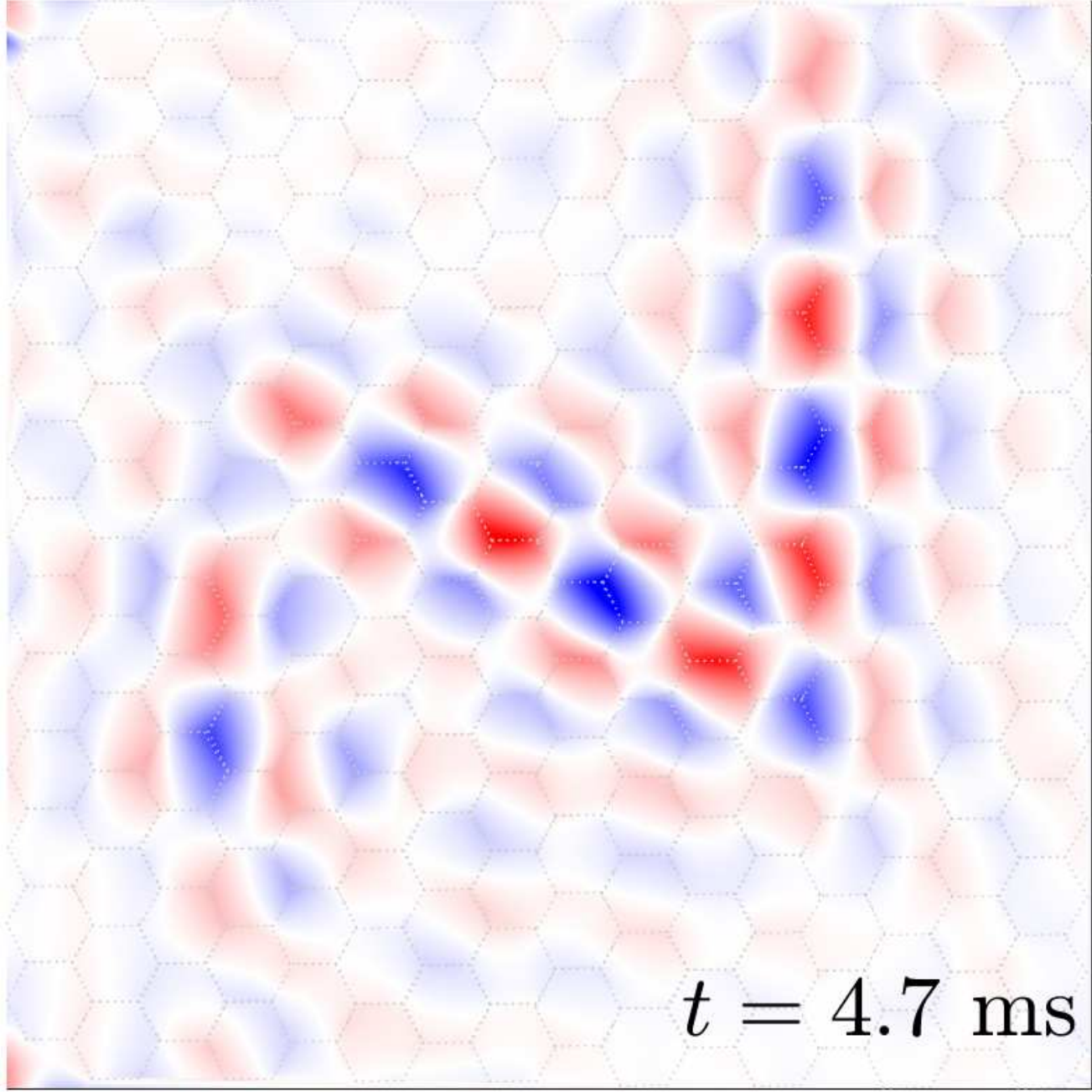}}
\caption{A nontrivial N-shaped interface (red dots indicate two masses attached at the sub-lattice sites and cyan dots denote locations where no masses are added) with $\gamma=(-)1$ on the left(right) (a). Snapshots of measured wave motion at two instants of time. Excitation is a 11 cycles tone burst at $3$ kHz (b).}
	\label{Fig.Waveguide2}
\end{figure}

A second example considers an N-shaped interface with segments parallel to the lattice vectors (Fig.~\ref{Fig.Waveguide2}a). The objective is to observe the behavior of the wave in the presence of 120\textdegree$\,$ corners along the interface. The cylindrical masses are attached so that the unit cells to the left top of the interface have $\gamma=1$ and those to the right bottom have $\gamma=-1$. The results in Figs. \ref{Fig.Waveguide2}b and \ref{Fig.Waveguide2}c show the propagation of the wave along the N-shaped topological interface, and a limited propagation into the bulk. Furthermore, the wave manages the 120$^o$ turn, illustrating the ability to change direction with limited backscattering. The contours are normalized by the displacement amplitude $1.5\times10^{-8}$ m. Movie files are provided in the supplementary material.

\begin{figure}[hbtp]
	\centering
\subfigure[]{\includegraphics[width=0.3\textwidth]{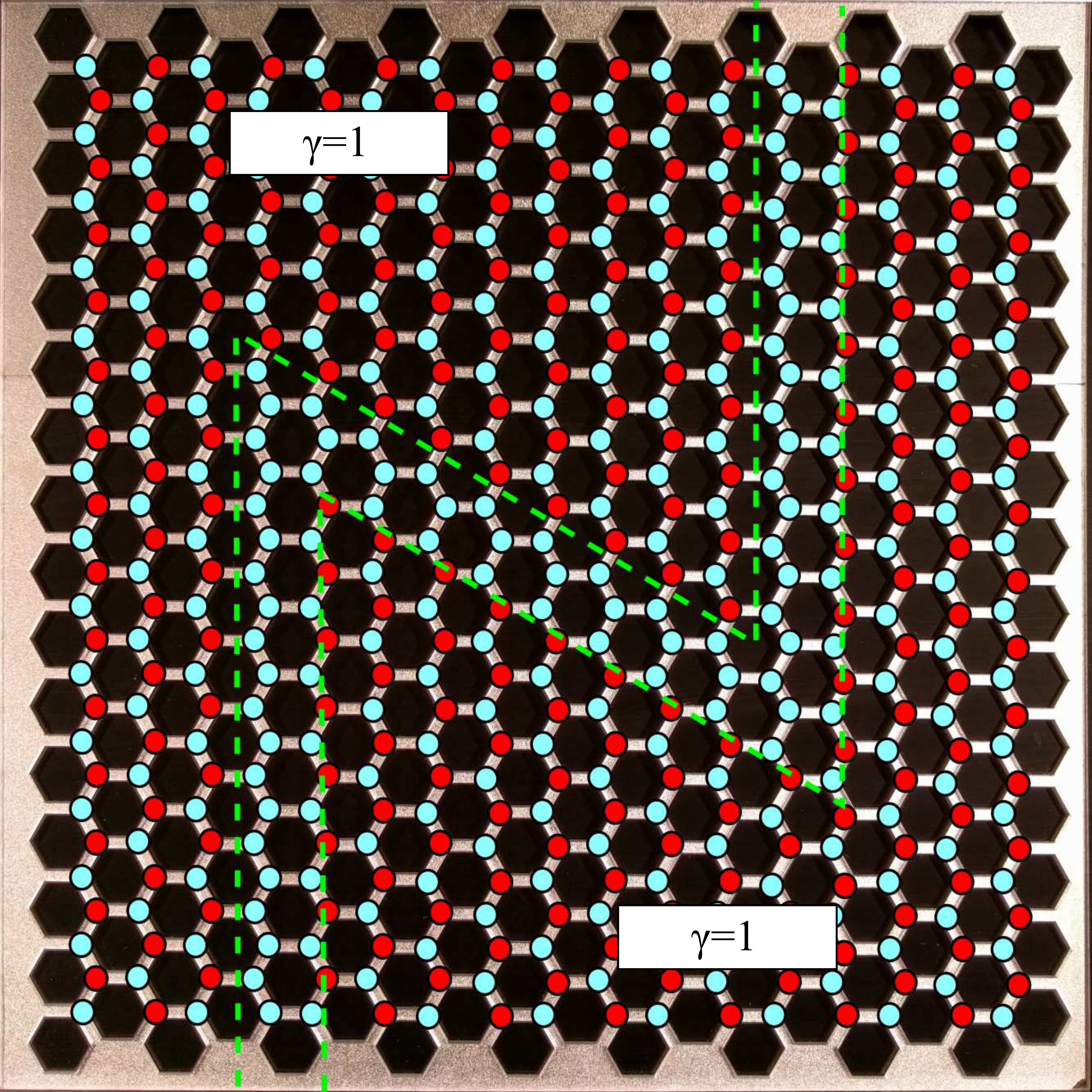}}\\
\subfigure[]{\includegraphics[width=0.3\textwidth]{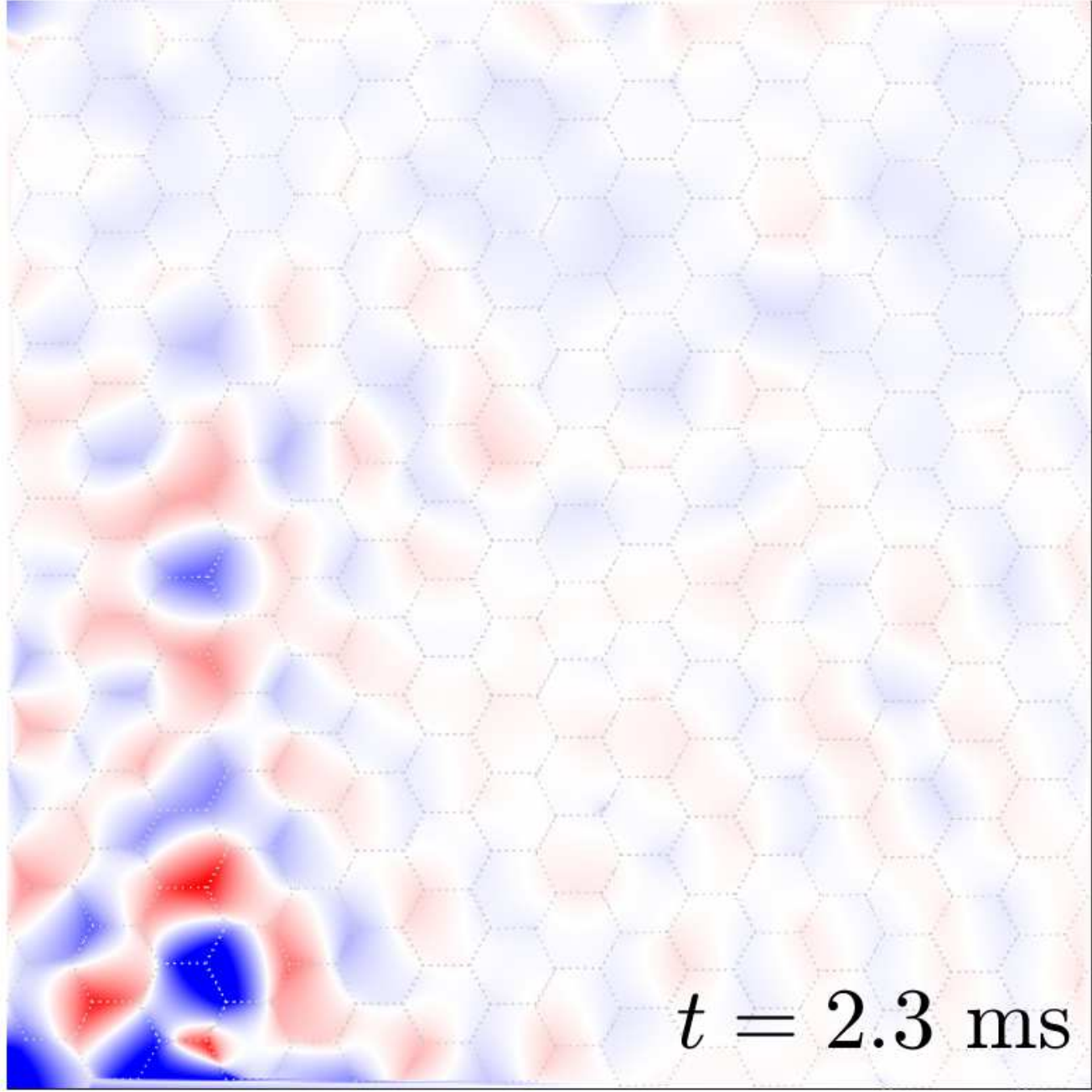}}
\subfigure[]{\includegraphics[width=0.3\textwidth]{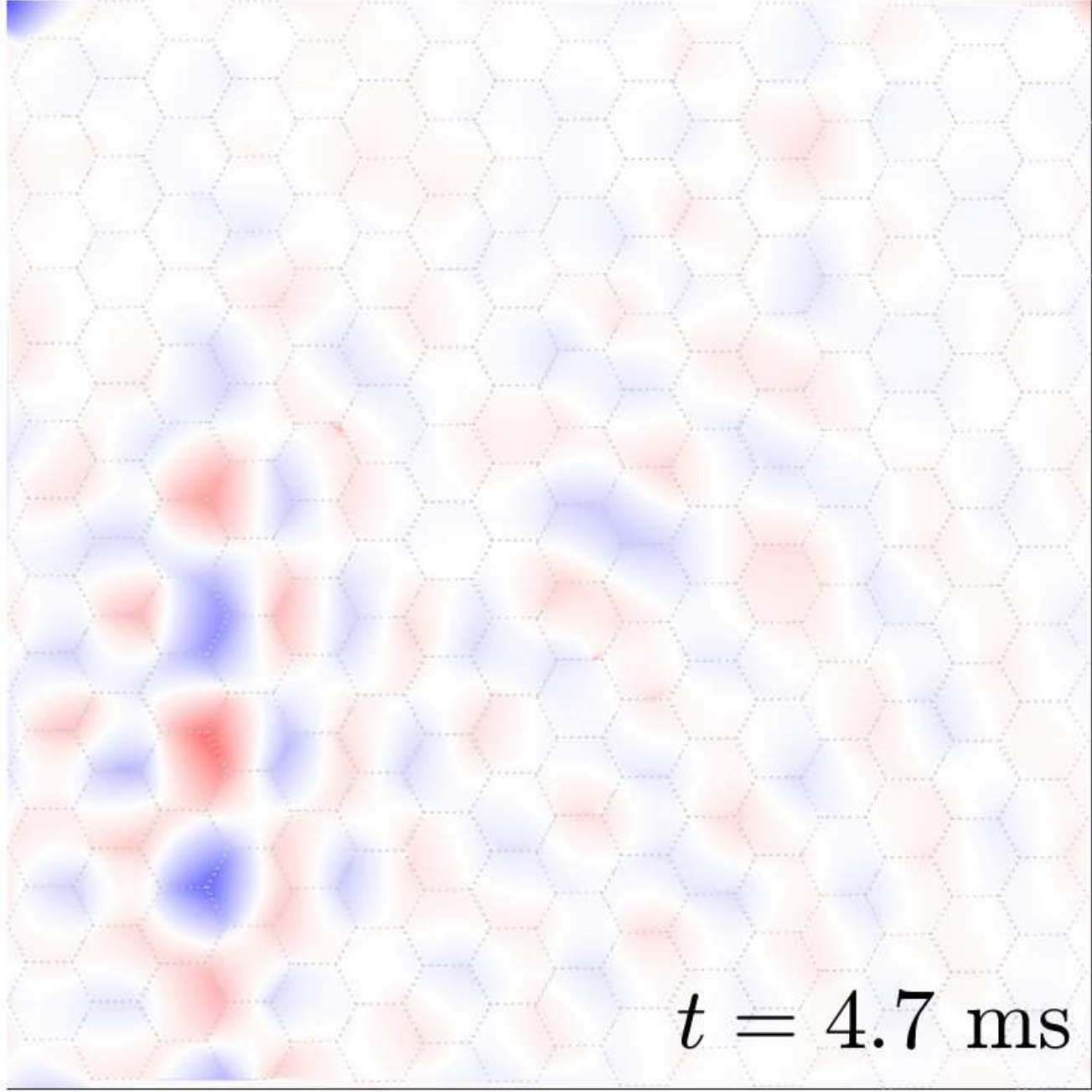}}
\caption{N-shaped trivial interface (red dots indicate two masses attached at the sub-lattice sites and cyan dots denote locations where no masses are added) with $\gamma=1$ everywhere (a). Snapshots of measured wave motion at two instants of time. Excitation is a 11 cycles tone burst at $3$ kHz (b).}
\label{Fig.Waveguide3}
\end{figure}

To illustrate that the waveguide happens due to the nontrivial topological nature of the interface modes, we compare an contrast its performance with a waveguide having a trivial (non-topological) interface. Masses are attached so that the unit cells at both sides of this trivial N-shaped interface have $\gamma=1$. This is achieved by tracing a N-interface within one lattice type, in this case corresponding to $\gamma=1$, simply by removing a line of masses, as illustrated in Fig.~\ref{Fig.Waveguide3}a. The resulting propagation is shown in Figs. \ref{Fig.Waveguide3}b and \ref{Fig.Waveguide3}c, where one clearly observes the limited ability of the wave to enter and propagate along the interface. The induced motion of the lattice appears to remain localized in the vicinity of the excitation point, and eventually decaying as a result of material dissipation. The contours are also normalized by the displacement amplitude $1.5\times10^{-8}$ m. Comparing the amplitude of transmitted waves, we conclude that the amount of energy traveling through the interface is much lower in this case. The topological interface is thus shown to demonstrate superior properties as a waveguide, since the TPEWs are immune to the width of the interface and the presence of sharp turns.

\section{Conclusion}\label{Sec.Concl}

Control of mechanical waves is applicable in many technological fields of interest, including detection, energy harvesting and telecommunications. This study demonstrates the existence of interface modes within the bandgap of a two-dimensional elastic hexagonal lattice and the propagation of TPEWs exploiting a mechanical analogue of the quantum valley Hall effect. This phenomenon allows creating a simple and robust waveguide for elastic waves in a wide band of frequencies. Guided by studies on conceptual lattices and numerical simulations, experiments are conducted to predict the dispersion properties of the considered hexagonal lattices and to explore the existence of TPEWs along predefined interfaces. The difference in propagation along non trivial interfaces is also illustrated through an experiment that reveals the fundamental difference of modes of propagation endowed with topological protection from those that are obtained by introducing a line defect in an otherwise periodic assembly. The experimental configurations illustrated herein are suitable for potential implementation of the concept to phononic systems and structural components, and could be further utilized to investigate the sensitivity of these configurations to a variety of defect and interface configurations.

\bibliography{papers}

\end{document}